\documentclass{IEEEtran}
\usepackage{amsthm}
\usepackage{bbm}
\usepackage{amssymb}
\usepackage{tabstackengine} 
\usepackage[english]{babel}
\newtheorem{theorem}{Theorem}
\newtheorem{corollary}[theorem]{Corollary}
\newtheorem{lemma}[theorem]{Lemma}
\usepackage{tikz}
\usepackage{caption}
\usepackage{subcaption}
\theoremstyle{definition}
 \usepackage{balance}
\newtheorem{definition}{Definition}
\newcommand{\figref}[1]{fig. \ref{#1}}
\usepackage{etoolbox}
\usepackage[]{nomencl}   
    \makenomenclature

\newcommand{\Drawlegend}[6]{
  \begin{tikzpicture}[baseline=-3pt]
    \draw[line width=#1, color=#3, #4] (0,0) -- (#2,0);
    \fill[color=#3] (#2/2,0) circle [radius=#5*#6/2];
  \end{tikzpicture}%
}

\providetoggle{nomsort}
\settoggle{nomsort}{false} 

\makeatletter
\iftoggle{nomsort}{%
    \let\old@@@nomenclature=\@@@nomenclature        
        \newcounter{@nomcount} \setcounter{@nomcount}{0}%
        \renewcommand\the@nomcount{\two@digits{\value{@nomcount}}}
        \def\@@@nomenclature[#1]#2#3{
          \addtocounter{@nomcount}{1}%
        \def\@tempa{#2}\def\@tempb{#3}%
          \protected@write\@nomenclaturefile{}%
          {\string\nomenclatureentry{\the@nomcount\nom@verb\@tempa @[{\nom@verb\@tempa}]%
          \begingroup\nom@verb\@tempb\protect\nomeqref{\theequation}%
          |nompageref}{\thepage}}%
          \endgroup
          \@esphack}%
      }{}
\makeatother

\renewcommand\nomgroup[1]{%
  \item[\itshape
  \ifstrequal{#1}{A}{Sets}
  {
    \ifstrequal{#1}{B}{Indices}
    {
      \ifstrequal{#1}{C}{Parameters}
      {
        \ifstrequal{#1}{D}{Variables}
        }
    }
    }]}

\usepackage[cmex10]{amsmath}

\makeatletter
\let\old@ps@headings\ps@headings
\let\old@ps@IEEEtitlepagestyle\ps@IEEEtitlepagestyle
\def\psccfooter#1{%
    \def\ps@headings{%
        \old@ps@headings%
        \def\@oddfoot{\strut\hfill#1\hfill\strut}%
        \def\@evenfoot{\strut\hfill#1\hfill\strut}%
    }%
    \def\ps@IEEEtitlepagestyle{%
        \old@ps@IEEEtitlepagestyle%
        \def\@oddfoot{\strut\hfill#1\hfill\strut}%
        \def\@evenfoot{\strut\hfill#1\hfill\strut}%
    }%
    \ps@headings%
}
\makeatother

\begin{document}

\title{Efficient Quantification and Representation of Aggregate Flexibility in Electric Vehicles}
%

\author{
\IEEEauthorblockN{Nanda Kishor Panda, Simon H. Tindemans }
\IEEEauthorblockA{Department of Electrical Sustainable Energy\\
Delft University of Technology, Delft, The Netherlands}
\{n.k.panda, s.h.tindemans\}@tudelft.nl
}

\maketitle
\begin{abstract}
Aggregation is crucial to the effective use of flexibility, especially in the case of electric vehicles (EVs) because of their limited individual battery sizes and large aggregate impact. This research proposes a novel method to quantify and represent the aggregate charging flexibility of EV fleets within a fixed flexibility request window. These windows can be chosen based on relevant network operator needs, such as evening congestion periods. The proposed representation is independent of the number of assets but scales only with the number of discrete time steps in the chosen window. The representation involves $2T$ parameters, with T being the number of consecutive time steps in the window. The feasibility of aggregate power signals can be checked using $2T$ constraints and optimized using $2(2^T-1)$ constraints, both exactly capturing the flexibility region. Using a request window eliminates uncertainty related to EV arrival and departure times outside the window. We present the necessary theoretical framework for our proposed methods and outline steps for transitioning between representations. Additionally, we compare the computational efficiency of the proposed method with the common direct aggregation method, where individual EV constraints are concatenated.
\end{abstract}
\thanksto{\noindent The research was supported by the ROBUST project, which received funding from the MOOI subsidy programme under grant agreement MOOI32014 by the Netherlands Ministry of Economic Affairs and Climate Policy and the Ministry of the Interior and Kingdom Relations, executed by the Netherlands Enterprise Agency. The authors would like to thank ROBUST consortium partners for fruitful discussions during the preparation of this paper.}


\section{Introduction}
Integrating Electric Vehicles (EVs) into the power network presents challenges and opportunities. Unregulated EV charging can strain the grid, but smart control of these processes can benefit the power networks~\cite{Muratori2018}. EV batteries offer flexibility as they are often available at charging stations for longer periods than the time required to charge them, creating a buffer of energy storage. This flexibility can be leveraged by adjusting charging power levels, delaying charging, or enabling bi-directional power flow to the grid~\cite{Smart2015}.\par
While leveraging the flexibility of a single EV is straightforward, its potential to support the grid is limited due to its smaller battery capacities. To effectively deploy EV flexibility at a large scale, it is necessary to aggregate the flexibility of individual EVs, taking into account their operational and technical constraints~\cite{taha2023efficient}. However, the aggregate control of EV fleets is a complex process that necessitates appropriate mathematical models, extensive calculations, advanced ICT infrastructure, and upgraded charging facilities and poses challenges, particularly in terms of scalability and accuracy~\cite{20.500.11850/202028}. This paper looks at the challenges associated with efficient and scalable aggregation of EV flexibility. \par

Aggregation of multiple flexible assets can be done using bottom-up or top-down approaches. Bottom-up approaches start from the properties of individual assets, which are combined to estimate the flexibility of the overall system. A common example is the \emph{direct aggregation} of EV flexibility, where constraints for individual EVs are concatenated to determine the overall system flexibility. 

This approach is frequently employed in classical EV scheduling problems, where charging schedules are optimized by one or more aggregators~\cite{XU2014582,8783654}. In many cases, direct aggregation can include other operational constraints, for instance, the power flow of the network~\cite{verbist2023impact}. 

One of the primary challenges of direct aggregation is the increase of model size with the number of assets. This rapidly results in a large computational burden; even when advanced solvers are used, making real-time utilization difficult. 

Another challenge lies in the lack of a concise representation for efficient information sharing and decentralized optimization.\par

Applying \emph{set-based aggregation} can effectively address many of the challenges associated with direct aggregation. In this approach, individual asset constraints are initially transformed into quantifiable parameters or metrics that can capture their unique flexibility characteristics. These parameters are then aggregated to determine the overall system's flexibility. Using aggregate parameters facilitates a more efficient representation; for instance, representing the feasible flexibility using convex polytopes can aid aggregators in participating in flexibility markets. This has the potential to enhance scalability and simplify decentralized decision-making. In the context of markets, it is imperative that such flexibility representations align with the market design~\cite{taha2023efficient}.\par

The exact aggregation of individual feasible flexibility sets (i.e., achievable power consumption patterns) is  
known as the Minkowski sum or simply M-sum. However, computing the M-sum is challenging and, in general, an NP-hard problem, which may become intractable in practice~\cite{Tiwary_2008}. Researchers have attempted to address this issue through various approximation methods, such as inner and outer approximation techniques~\cite{taha2023efficient, homothet_zhao,7403253, Barot2017}. For instance, in \cite{Wu2020}, the authors introduced a virtual battery model to aggregate flexibility from thermostatically controlled loads in buildings for providing grid services by summing up individual asset parameters to obtain the aggregate approximation, an outer approximation of the feasible flexibility. At a larger scale, power system units such as generators can be aggregated by adding their individual flexibility metrics, such as energy, power and ramp rates~\cite{ulbig2015analyzing}. However, such direct summation of individual flexibility constraints results in outer approximations that are overly optimistic regarding aggregate flexibility. To avoid this, inner approximations based on zonotopes (\cite{7403253}), inner volume maximisation (\cite{taha2023efficient}), and homothets (\cite{homothet_zhao}) have been proposed. Although inner approximations will not lead to infeasible operating points, they may grossly underestimate flexibility, potentially causing economic losses. Striking a balance between accuracy and computational complexity is crucial in such scenarios. 

Top-down approaches such as \cite{Brinkel2022} directly approximate EV fleet flexibility. Data-driven approximation methods have emerged that directly identify parameters of aggregate EV load models and simulate EV charging demand under diverse electricity market scenarios~\cite{9069450WANG}. However, such data-driven approaches may suffer from computational complexity, limited scalability, and high specificity to particular cases. \par

Exact set-based aggregation methods have been developed for the special case of discharging (or charging) of heterogeneous stationary batteries in \cite{9772074_ep, cruise2019optimal, Zachary2021}. The E-p and E-t representations proposed in  \cite{9772074_ep} and \cite{cruise2019optimal} respectively present a continuous-time exact representation of feasible flexibility. However, unlike our proposed methodology, which considers discrete timesteps, these methods don't include minimum charge requirements of batteries. This makes them less suitable for application to EV smart charging scenarios, where minimum charge requirements are often enforced. Further, in their continuous-time form, these representations are not easy to integrate into an optimization framework. Concurrently with the work presented in this paper, the authors of 
 \cite{mukhiexact} have developed an exact characterization of flexibility in a population of EVs that is independent of fleet size, for vehicles having the same arrival and departure time. However, it strictly enforces full charging by the end of the interval, which reduces the application scope. 

This paper proposes a novel method for aggregating the flexibility of EVs across a set of discrete time intervals, which can be used for efficient scheduling of EV fleet power consumption. 
The method, based on set-based aggregation concepts, offers scalability and independence from the size of the EV population. The proposed representations incorporate both maximum \emph{and} minimum charging requirements, which is essential for practical fleet management. Secondly, we show different representations (including H- and V-representations) for the aggregate EV flexibility and their applicability to various use cases. Further, we show that the proposed method can achieve considerable computational gain in terms of time and memory when scheduling EVs compared with direct aggregation methods. \par

\begin{figure}
    \centering
    \includegraphics[width=\linewidth]{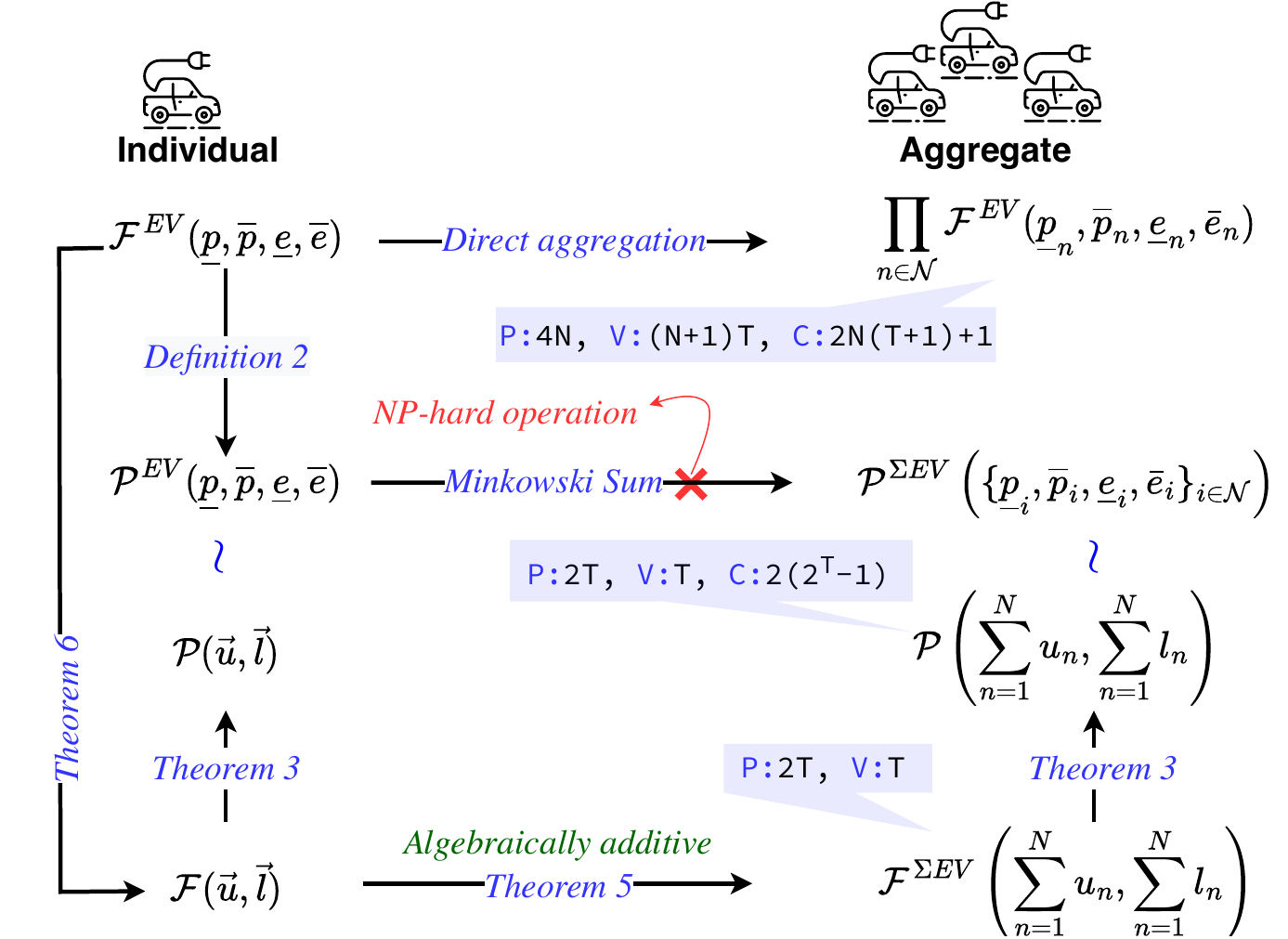}
    \caption{Different approaches for aggregating flexibility in EVs. Computational complexity is indicated by the number of parameters (P), variables (V) and constraints (C), where N is the number of vehicles and T is the number of time steps}
    \label{fig:different_methods}
\end{figure}

\section{Problem definition}\label{sec_background}

\subsection{Notation}
Matrices are denoted in upper case ($M$), vectors by arrows ($\Vec{v}$), and their components by subscripts ($M_{ij}, v_i$) respectively. A subscript referring to a specific EV may be added to a vector ($\Vec{v}_n$). Element-wise comparison between vectors is indicated by $\prec$, $\preceq$, $\succ$ and $\succeq$. We use EV as shorthand for battery EV (BEV).

\subsection{Single EV flexibility}

We consider the smart charging of an EV during a flexibility window $\mathcal{T}$ consisting of $T$ discrete time intervals indexed by $t=1,\dots,T$. Each interval is of duration $\Delta \text{t}$, and power consumption is assumed to be constant within the interval. The charge requirements and flexibility of each EV are determined by four parameters: the lower and upper charging rate limits $(\underline{p}, \overline{p})$ and the minimum and maximum energy requirements $(\underline{e}, \overline{e}$). These parameters are assumed to be internally consistent, e.g., $\underline{e} \ge T \underline{p}$, and discharging is not permitted ($\underline{p}\ge 0$). Together, the parameters constrain the power $\vec{p} \in \mathbb{R}^T$ with which the EV can charge as follows:
\begin{align}
    \underline{p} & \le p_t  \le \overline{p} && \forall t \in \mathcal{T} \label{eq:plimit} \\
    \underline{e} & \le \Delta \text{t} \sum_{t \in \mathcal{T}} p_t  \le \overline{e} && \forall t \in \mathcal{T} \label{eq:elimit}
\end{align}
It is straightforward to verify that this is a convex set.
\begin{definition}[Feasible EV flexibility]\label{def_1} 
The feasible flexibility $\mathcal{F}^{EV} \equiv \mathcal{F}^{EV}(\underline{p},\overline{p}, \underline{e}, \overline{e})$ is the set of permissible power levels 
\begin{equation}
    \mathcal{F}^{EV}(\underline{p},\overline{p}, \underline{e}, \overline{e}) := \{ \vec{p} \in \mathbb{R}^T |\: \vec{p} \text{ subject to \eqref{eq:plimit}, \eqref{eq:elimit}} \} \label{eq:FEV-definition}
\end{equation}
\end{definition}
\begin{definition}[Polytope representation of feasible EV flexibility]
The feasible flexibility of an EV $\left(\mathcal{F}^{EV}\right)$ can also be written in the form of a polytope ($\mathcal{P}^{EV} \equiv \mathcal{P}^{EV}(\underline{p},\overline{p}, \underline{e}, \overline{e})$) whose H-representation (c.f. \cite{sigl2023mrepresentation}) is given as:
\begin{align}
   \mathcal{P}^{EV} :=\left\{ \vec{p} \in \mathbb{R}^T \Bigg| \begin{bmatrix}
       \:\:\: \mathbb{I}\\
        -\mathbb{I}\\
        \:\:\: (1)^\top\Delta \text{t}\\
        - (1)^\top\Delta \text{t}
    \end{bmatrix}\vec{p} \preceq \begin{bmatrix}
        \:\:\:(1)\overline{p}\\
        -(1)\underline{p}\\
        \:\:\:(1)\overline{e}\\
        -(1)\underline{e}
    \end{bmatrix}  \right\}, \label{eq:FEV polytope-definition}
\end{align}
where $(1)$ is a column vector of length $T$ with elements equal to 1.
\end{definition}

\subsection{Aggregate flexibility}

Having defined the charging flexibility of a single EV, we consider a population of EVs indexed by $n \in \mathcal{N} = \{1,\ldots,N\}$. We assume all EVs under consideration are connected during the window $\mathcal{T}$. In reality, vehicles are connected to and disconnected from chargers at different times during the day. This idealised model representation accurately represents various relevant use cases:
\begin{itemize}
    \item The period $\mathcal{T}$ is a relatively short flexibility request window. Only EVs that are completely available during this window are assumed to contribute to system flexibility. The minimum and maximum charge levels of the EV may be adjusted to reflect the amount of charge that must/can be stored during the window.
    \item Large fleets of EVs, such as commercial delivery vehicles or buses, spend the night hours in a depot connected to a charger. The flexibility request window where (nearly) all vehicles are connected can be substantial in this case.
    \item For very large fleets, we may consider only the flexibility of all vehicles that connect around time $t_1$ and disconnect around time $t_2$, and construct different flexibility representations for all relevant combinations of $t_1, t_2$.
\end{itemize}

To utilize the flexibility of all EVs in $\mathcal{N}$ effectively, it is essential to understand the operational and technical constraints of the aggregate system. The charging flexibility of each individual EV $n$ is given by \eqref{eq:FEV-definition}, with vehicle-specific parameters $(\underline{p}_n, \overline{p}_n, \underline{e}_n, \overline{e}_n)$. The remainder of this paper is concerned with the question of how to represent the feasible flexibility of a \emph{fleet} of EVs, i.e. the set of feasible aggregate power signals $\vec{P}$, where
\begin{equation}
    \vec{P} = \sum_{n \in \mathcal{N}} \vec{p}_n. \label{eq:sumpower}
\end{equation}
We denote the aggregate flexibility by $\mathcal{F}^{\Sigma EV} = \{ \vec{P} \}$, with constraints derived from the elementary EV flexibilities. 

Two common approaches for determining $\mathcal{F}^{\Sigma EV}$ exist. The first is by \emph{direct aggregation} (\cite{XU2014582,8783654,verbist2023impact}), where \eqref{eq:sumpower} is directly embedded in the problem, alongside \eqref{eq:plimit}-\eqref{eq:elimit} for each EV. This effectively constructs $\mathcal{F}^{\Sigma EV}$ as a $NT$-dimensional polytope projection to $T$ dimensions. A downside of this approach is that the complexity scales linearly with $N$: this representation has $4N$ parameters, $(N+1)T$ variables, $2N(T+1)$ inequalities, and equality \eqref{eq:sumpower}. For large $N$, this becomes highly inefficient and increases memory and processing speed requirements.

A second approach is the direct representation of $\mathcal{F}^{\Sigma EV} $ as a polytope in $\mathbb{R}^T$, a geometric object that is bounded by a finite number of facets that are generated by intersecting hyperplanes. It can provide a concise representation of feasible flexibility by capturing the feasible points that lie within the boundaries. Each hyperplane corresponds to a different constraint, allowing for a clear delineation of the feasible region. Polytopes are valuable as they enable efficient optimization, constraint modelling, and analysis of feasible solutions. The \emph{set-based aggregation} into a polytope is written as~\cite{Barot_2017}:
\begin{equation}
    \mathcal{F}^{\Sigma EV} = \mathcal{F}^{EV}_1 \oplus \mathcal{F}^{EV}_2 \oplus \ldots \oplus \mathcal{F}^{EV}_n, \label{eq:msumFEV}
\end{equation}
where the $\oplus$ operator represents Minkowski summation.
\begin{definition}[Minkowski summation] \label{def:Minkowski}
The Minkowski sum of two polytopes in $\mathbb{R}^T$ defined by the two sets $\mathcal{F}_1 \subseteq \mathbb{R}^T$ and $\mathcal{F}_2 \subseteq \mathbb{R}^T$, is defined as
    \begin{align}
        \mathcal{F}_1 \oplus \mathcal{F}_2 = \{\Vec{p}_1 + \Vec{p}_2 \: | \: \Vec{p}_1\in\mathcal{F}_1,\: \Vec{p}_2\in \mathcal{F}_2\} \label{eq:msum-definition}
    \end{align}
\end{definition}
It follows from \eqref{eq:msum-definition} that Minkowski summation is associative and commutative so that the aggregation can be done in any order. Moreover, as \eqref{eq:FEV-definition} is convex, so is \eqref{eq:msumFEV}. By definition, this procedure results in a $T$-dimensional representation of $\mathcal{F}^{\Sigma EV}$ ($T$ variables), but the process of determining the constraints is NP-hard in general, and direct approaches to compute \eqref{eq:msumFEV} \emph{without approximations} have so far not been successful~\cite{Tiwary_2008,7403253}. 

Fig.~\ref{fig:different_methods} shows both approaches to aggregating flexibility alongside the approaches that will be developed in the remainder of this paper, with references to relevant theorems and definitions.

\subsection{Motivating example}\label{section: motivating example}
Let us consider two EVs that are connected for a duration of three-time steps of 1~hr. The first EV has a maximum charge rate of 20~kW, a minimum charge rate of 0~kW, and an energy requirement of 15--25~kWh. If desired, it can satisfy its minimum charge requirements in a single time step. The second has permissible charge rates between 5 kW and 10 kW and requires a total energy of 20--30 kWh; it needs a minimum of two-time steps to satisfy its charge requirements.\par
A representation as a `simple virtual battery' that is obtained by the addition of the parameters of both EVs (as proposed in ~\cite{ulbig2015analyzing,Brinkel2022}) results in limits on the instantaneous power consumption of 5--30 kW and energy consumption of 35--55~kWh. This representation is overly optimistic: at the low end of power consumption, the aggregate representation would suggest that $\vec{p}=(5,30,0)^T~kW$ is a valid solution, but this would force EV \#2 to charge at an unattainable rate during the second interval to meet its minimum charge requirement. At the high end of power consumption, the solution $\vec{p}=(25,30,0)^T~kW$ fails because it either violates the upper energy limit of EV \#1 or the upper power limit of EV \#2.\par

The feasible flexibility of an individual EV as defined in \eqref{eq:FEV-definition} has three generic properties:
\begin{enumerate}
    \item It is restricted to positive values ($\mathcal{F}^{EV} \subseteq \mathbb{R}^{T}$).
    \item It considers discrete time intervals within which the power consumption is constant.
    \item Permutations of feasible charging patterns $\vec{p}$ are also feasible. This also implies that the EV must be connected during all $T$ time steps.
\end{enumerate}
In this section, we develop the concept of UL-flexibility, a \emph{discrete-time permutable flexibility representation} that is built on these properties.


\subsection{Definition}

In analogy with a view of vector $\vec{v}$ as function a function $v(i)\equiv v_i$, we define the following.

\begin{definition}[Convex/concave vectors]
A vector $\vec{v} \in \mathbb{R}^T$ is convex iff $v_j \ge (v_i + v_k)/2$ or concave iff $v_j \le (v_i + v_k)/2$, for any $(i,j,k) \in \{1,\ldots,T\}^3$ with $i<j<k$.
\end{definition}

\begin{definition}[Zero-extended vectors]
For any vector $\vec{v} \in \mathbb{R}^{T}$ we define the zero-extended vector ${}^0\vec{v}$ as $(0, v_1, \ldots, v_T)^T \in \mathbb{R}^{T+1}$. 
\end{definition}

Using these auxiliary concepts, we define the UL-flexibility representation for an energy consuming flexible asset as:
\begin{definition}[UL-flexibility]  \label{def:ULflex}
The UL-flexibility $\mathcal{F}(\vec{u},\vec{l})$, with $\vec{u},\vec{l} \in \mathbb{R}^T$, is given by the set of all signals $\vec{p} \in \mathbb{R}^{T}$ for which, for all $k \in \{1,\dots,T\}$:
\begin{enumerate}
    \item  The energy consumed in any $k$ intervals does not exceed the upper bound $u_k$;
    \item  The energy consumed in any $k$ intervals does not drop below the lower bound $l_k$.
\end{enumerate}
The vectors $\vec{u}$ and $\vec{l}$ must satisfy the following properties:
    \begin{enumerate}
        \item[i)] The vector ${}^0\vec{u}$ is concave, and element-wise non-negative and increasing.
        \item[ii)] The vector ${}^0\vec{l}$ is convex, and element-wise non-negative and increasing.
        \item[iii)] The vector ${}^0\vec{u} +  R \cdot {}^0\vec{l}$, with the order-reversing matrix $R\in \mathbb{R}^{T + 1 \times T + 1}$, is element-wise increasing.
    \end{enumerate}
\end{definition}

signal.

\begin{lemma}[Ordered UL representation]
The UL-flexibility for vectors $\vec{u}, \vec{l}$ can be expressed as 
\begin{equation}
        \mathcal{F}(\Vec{u},\vec{l})=
         \left\{\Vec{p}\in \mathbb{R}^{+^T} \;\middle\vert\;
   \begin{array}{r@{}l}
   \vec{\tilde{p}} \; & := \text{descending}(\vec{p}) \\
   \Delta \text{t} \mathbbm{1}^L \Vec{\Tilde{p}} \; & \preceq \vec{u} \\
   \Delta \text{t} \mathbbm{1}^L R \Vec{\Tilde{p}} \;  & \succeq \vec{l} 
   \end{array}
   \right\}. \label{eq:ULordered}
\end{equation}
Here, $\vec{\tilde{p}}$ is the component-wise descending permutation of $\vec{p}$, $\mathbbm{1}^L \in \mathbb{R}^{T\times T}$ is the lower triangular matrix where all lower triangular elements are 1 and $R\in \mathbb{R}^{T\times T}$ is the order reversing matrix, where the 1 elements reside on the antidiagonal and all other elements are zero.
\end{lemma}
\begin{proof}
Definition~\ref{def:ULflex} states the constraints in an inherently permutation invariant manner (``any $k$ intervals''), so that we may restrict ourselves to the decreasing signal $\vec{\tilde{p}} := \text{descending}(\vec{p})$. The largest energy consumption in $k$ intervals is then given by the partial sum over the $k$ \emph{first} intervals: $\Delta \text{t} \sum_{j=1}^k \tilde{p}_j$. Imposing the energy upper bound for all $k$ results in $\Delta \text{t} \mathbbm{1}^L \Vec{\Tilde{p}} \preceq \vec{u}$. Similarly, the partial sum over the \emph{last} $k$ values of the ordered vector $\vec{\tilde{p}}$ reflects the lowest energy consumption during $k$ intervals, resulting in the final constraint $\Delta \text{t} \mathbbm{1}^L R \Vec{\Tilde{p}} \succeq \vec{l}$.
\end{proof}

\begin{figure}
    \centering
    \includegraphics[width=\linewidth]{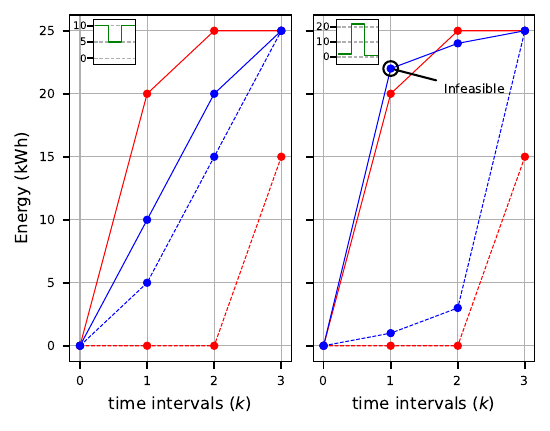}
    \caption{Graphical check of feasibility using ordered UL-flexibility. The UL values plotted here correspond to EV 1 as explained in section \ref{section: motivating example}. $\Vec{u}$ and $\Vec{l}$ for the EV is represented by \protect\Drawlegend{0.5pt}{0.5cm}{red}{solid}{1}{4pt}\protect \hspace{0.1 cm} and \protect\Drawlegend{0.5pt}{0.5cm}{red}{dashed}{1}{4pt}\protect \hspace{0.1 cm} respectively. The feasibility of the reference signals (\protect\Drawlegend{0.5pt}{0.5cm}{green}{solid}{0}{4pt}\protect) is checked by comparing the $\sum \textrm{descending}(\Vec{p})$ (\protect\Drawlegend{0.5pt}{0.5cm}{blue}{solid}{1}{4pt}\protect) and $ \sum \textrm{ascending}(\Vec{p})$ (\protect\Drawlegend{0.5pt}{0.5cm}{blue}{dashed}{1}{4pt}\protect) by their respective upper and lower limits.}
    \label{fig:graphical_check}
\end{figure}

The ordered representation presented in \eqref{eq:ULordered} can come in handy to check the feasibility of a discrete power signal graphically. For an asset with flexibility parameterised by $\vec{u}$ and $\vec{l}$, we can arrange the reference power signal in ascending and descending order. For a signal to be feasible (cf.~\figref{fig:graphical_check}), the integral of the ascending power signal should always be greater or equal to the lower energy bounds ($\vec{l}$), and the integral of the descending power signal should be smaller or equal than the upper energy bound ($\vec{u}$). Fig.~\ref{fig:graphical_check} shows an examples for $\vec{u}, \vec{l}$ given by \eqref{eq:ul_example1} , corresponding to the parameters of EV 1 in section:\ref{section: motivating example}. The reference power signal [10,5,10] kW (left plot)  is feasible as it satisfies \ref{eq:ULordered}, whereas the reference signal [2,22,11] kW is infeasible as it violates $u_1$ (power at t=2 exceeds the maximum power). 

For the simpler case with $\vec{l}=0$, this graphical approach is equivalent to the E-t diagram presented in \cite{Zachary2021} and (in different coordinates) the E-p diagram \cite{9772074_ep}, when applied to piecewise constant power signals.

\begin{lemma}[Properties of UL-flexibility]
A valid UL-flexibility set $\mathcal{F}(\vec{u},\vec{l})$ satisfies the following properties:
\begin{enumerate}
        \item[i)] The pair $\vec{u}, \vec{l}$ satisfy $\vec{l} \preceq \vec{u}$. 
        \item[ii)] $\mathcal{F}(\vec{u},\vec{l})$ is non-empty.
\end{enumerate}
\end{lemma}
\begin{proof}
Property (i) follows directly from the fact that $l_T \le u_T$ (implied by (iii) in Definition~\ref{def:ULflex}), combined with the knowledge that ${}^0\vec{u}$ is concave and ${}^0\vec{l}$ is convex. 
To prove (ii), we can define the constant power signal $\vec{p}^{\,c} = (c,\dots, c)^T$, with $c=u_T / (T \Delta \text{t})$. Concavity of ${}^0\vec{u}$ and convexity of ${}^0\vec{l}$, together with \eqref{eq:ULordered} show that this is a feasible signal. Therefore, the set $\mathcal{F}(\vec{u},\vec{l})$ is non-empty.
\end{proof}

\subsection{Polytope representation}

\begin{theorem}[Polytope, H-representation]
The UL-flexibility set $\mathcal{F}(\vec{u},\vec{l}) = \mathcal{P}(\vec{u},\vec{l})$, where the polytope can be represented in H-space as:

\begin{equation} \label{eq:polyrepr}
\mathcal{P}(\vec{u},\vec{l}) = \left\{ \vec{p} \in \mathbb{R}^T \Bigg| \begin{bmatrix}
    \:\:\: A\\
    -A
\end{bmatrix}\vec{p} \preceq 
\begin{bmatrix}
    \:\:\:B\vec{u}\\
    -B\vec{l}
\end{bmatrix},
 \right\}
 \end{equation}
with the $(2^T - 1)\times T $ matrices
\begin{equation}
 A = \left(  \begin{array}{c} C_1 \\ C_2 \\ \vdots \\ C_T \end{array}  \right), B = \left(  \begin{array}{cccc}
(1)_1 & (0)_1 &  \cdots &  (0)_1 \\
(0)_2 & (1)_2 &  \cdots &  (0)_2 \\
\vdots & \vdots &  \ddots&  \vdots \\
 (0)_T & (0)_T &  \cdots&  (1)_T \\
\end{array}   \right).\label{perm_a_b}
\end{equation}
Here, $(0)_i$ and $(1)_i$ represent column vectors of length $T \choose i$ with values $0$ and $1$, respectively, and $C_i$ is the ${T \choose i} \times T$ matrix with rows that contain all permutations of $i$ 1s and $T-i$ 0s.
\end{theorem}

\begin{proof}
The representation follows directly from Definition~\ref{def:ULflex}, by enumerating all permutations of $k$ time intervals and matching them with the relevant bounds $l_k$ and $u_k$.
\end{proof}

\begin{lemma}[Polytope, V-representation] \label{lem:spanning}
The polytope $\mathcal{P}(\vec{u},\vec{l})$ is spanned by the set of vertices
\begin{equation}
\mathcal{V} = \left\{ \Pi \vec{p}^{\,(k)} \in \mathbb{R}^T \big| \Pi \in \text{perm}(T), k \in \{ 0,\ldots,T\} \right\}, \label{eq:v_rep}
\end{equation}
where $\text{perm}(T)$ is the set of $T$-dimensional permutation matrices and $\vec{p}^{\,(k)}$ is the unique decreasing vector defined by equality of the $T$ constraints related to $\{u_1, \ldots, u_k\}$ and $\{l_1, \ldots, l_{T-k}\}$. 
\begin{equation} \label{eq:vertexdefinition}
    p^{\,(k)}_t = \left\{ \begin{array}{l@{}l}
   (u_t - u_{t-1})/\Delta \text{t}, & \qquad \text{for } t \le k, \\
   (l_{T-t+1} - l_{T-t})/\Delta \text{t}, & \qquad \text{for } t > k,
   \end{array} \right.
\end{equation}
where we define $u_0 = l_0 = 0$ for notational simplicity.
\end{lemma}

\begin{proof}
As a convex, closed polytope with a finite number of constraints, $\mathcal{P}(\vec{u},\vec{l})$ is spanned by a finite set of vertices. Each vertex is a feasible point, characterised by $T$ non-identical \emph{binding} constraints in \eqref{eq:polyrepr}~\cite{Grünbaum_Klee_Ziegler_2003}. Because of permutation invariance, we restrict ourselves to analysing vertices that correspond to decreasing power signals. The remaining vertices are generated by multiplication by $\Pi \in \text{perm}(T)$. For a decreasing power signal, the active constraints are indexed unambiguously by the lower and upper energy limits: $l_1,\ldots, l_T$ (with $l_k$ applying to the final $k$ time steps) and $u_1,\ldots, u_T$ (with $u_k$ applying to the first $k$ time steps). Identifying the vertices is equivalent to determining which of the $2T$ constraints are simultaneously active. 

Let us suppose that $u_k$ and $l_{T-k+i}$ are simultaneously active for any positive integer $i \le k$. Let us denote the energy consumed during the first $k-i$ intervals by $A$, the energy consumed during the next $i$ intervals by $B$ and the energy consumed during the final $T-k$ intervals by $C$. We then have $A+B=u_k$ (active constraint), but also $A \le u_{k-i}$ (definition of $u_{k-i}$). Therefore, $B\ge u_k - u_{k-i}$. We also have $B+C=l_{T-k+i}$ and $C\ge l_{T-k}$, so that $B \le l_{T-k+i} - l_{T-k}$. Combining these yields $u_k - u_{k-i} \le l_{T-k+i} - l_{T-k}$. However, property (iii) from Definition~\ref{def:ULflex} implies $u_k - u_{k-i} \ge l_{T-k+i} - l_{T-k}$ (opposite inequality). 

This yields two possibilities: (i) simultaneous activation of constraints $u_k$ and $l_{T-k+i}$ is impossible; (ii) the parameters satisfy the equality $u_k - u_{k-i} = l_{T-k+i} - l_{T-k}$. In the latter case, we also know that $A=u_{k-i}$, $B=u_k$, $C=l_{T-k}$, which results in $l_{T-k+i}$ being a redundant constraint. We conclude that $l_{T-k+i}$ is never (independently) active when $u_k$ is active. 

For a given vertex, let $k$ be the \emph{largest} index corresponding to an active constraint $u_k$, with $k=0$ if no upper energy limit constraints are active. Because we can disregard constraints $l_{T-k+i}$, for $i \in \{1,\ldots,k\}$, the only way to attain $T$ active constraints under this assumption is by considering the set $\{u_1, \ldots, u_k, l_1, \ldots, l_{T-k}\}$. This implies there are at most $T+1$ vertices that correspond with decreasing power signals, one for each $k$.

Next, we show that these $T+1$ power signals ($\vec{p}^{(k)}$, indexed by $k$) are positive, decreasing and uniquely defined by the constraints. Constraints $\{u_1, \ldots, u_k\}$ define the values $p_t^{(k)} = (u_t - u_{t-1})/\Delta \text{t}$ for $t \le k$, with $u_0=0$. Because ${}^0\vec{u}$ is concave and increasing, the sequence $p_{0:k}^{(k)}$ is positive and decreasing. Similarly, constraints $\{l_1, \ldots, l_{T-k}\}$ define $p_t^{(k)} = (l_{T-t+1} - l_{T-t})/\Delta \text{t}$ for $t > k$, with $l_0=0$. Because ${}^0\vec{l}$ is convex and increasing, the sequence $p_{k+1:T}^{(k)}$ is positive and decreasing.

For the whole power sequence to be positive and decreasing, it remains to be shown that $p_{k}^{(k)} \ge p_{k+1}^{(k)}$. Inserting \eqref{eq:vertexdefinition} yields the requirement
\begin{equation} \label{eq:proofconnectionstep}
    u_k - u_{k-1} \ge l_{T-k} - l_{T-k-1}.
\end{equation}
Property (iii) of Definition~\ref{def:ULflex} guarantees that $l_{T-k} - l_{T-k-1} \le u_{k+1} - u_k$. The concavity of ${}^0\vec{u}$ then implies \eqref{eq:proofconnectionstep}. Finally, Property (iii) of Definition~\ref{def:ULflex} also implies that the signal $\vec{p}^{(k)}$ cannot violate the inactive (or redundant) constraints $\{u_{k+1},\ldots, u_T \}$ and $\{ l_{T-k+1} ,\ldots, l_T \}$.

This concludes the proof: for every $k$ a vertex is uniquely defined by \eqref{eq:vertexdefinition}, and it corresponds to a valid, decreasing power signal.

\end{proof}

\subsection{Aggregation of UL-flexibility}

\begin{theorem}[Minkowski summation of UL-flexibility]\label{theorem:ul-aggregation}
Minkowski summation of feasible sets for UL-flexibility is additive in parameters, i.e., 
    \begin{align}
\mathcal{F}(\Vec{u}_1,\Vec{l}_1)\oplus \mathcal{F}(\Vec{u}_2,\Vec{l}_2) &= \mathcal{F}\left(\Vec{u}_1+\Vec{u}_2,\Vec{l}_1+\Vec{l}_2\right) \label{eq:Minkowski-Fduo}
    \\
    \intertext{and more generally,}
    \mathcal{F}(\Vec{u}_1,\Vec{l}_1)\oplus\cdots \oplus \mathcal{F}(\Vec{u}_n,\Vec{l}_n) &= \mathcal{F}\left(\sum_n\Vec{u}_n,\sum_n\Vec{l}_n\right). \label{eq:Minkowski-F} 
    \end{align}
\end{theorem}
\begin{proof}
To prove \eqref{eq:Minkowski-Fduo}, we first show that if $\vec{p}_1 \in \mathcal{F}(\vec{u}_1, \vec{l}_1) \equiv \mathcal{F}_1$ and $\vec{p}_2 \in \mathcal{F}(\vec{u}_2, \vec{l}_2) \equiv \mathcal{F}_2$, then $\vec{p}_1+\vec{p}_2 \in \mathcal{F}\left(\Vec{u}_1+\Vec{u}_2,\Vec{l}_1+\Vec{l}_2\right) \equiv \mathcal{F}_{12}$. This follows directly from \eqref{eq:polyrepr}. 

The reverse implication requires that 
\begin{equation} \label{eq:revrequirement}
    \forall \vec{p} \in \mathcal{F}_{1+2}: \exists (\vec{p}_1 \in \mathcal{F}_1, \vec{p}_2 \in \mathcal{F}_2), \vec{p}_1 + \vec{p}_2 = \vec{p}
\end{equation}
This follows from Lemma~\ref{lem:spanning}: UL-flexibility sets are convex and spanned by vertices $\mathcal{V}$. Therefore, if \eqref{eq:revrequirement} holds for $\vec{p}\in \mathcal{V}_{1+2}$, then it holds for all $\vec{p}\in \mathcal{F}_{1+2}$ by taking positive linear combinations of the vertices. Moreover, the set of vertices is defined by \eqref{eq:vertexdefinition}, which is linear in $\vec{u}$ and $\vec{l}$. Hence, there is a valid and unique solution to \eqref{eq:revrequirement} for each vertex. This completes the proof of \eqref{eq:Minkowski-Fduo}. Finally, repeated application of \eqref{eq:Minkowski-Fduo} proves \eqref{eq:Minkowski-F} for arbitrary (finite) $n$. 

\end{proof}

\section{Representing EV flexibility}

In this section, we will use the previously developed UL-flexibility to represent and aggregate EV flexibility efficiently.

\begin{theorem}[UL parameterization of EV]\label{theorem: EV's UL flexibility} The feasible flexibility of an EV satisfying \eqref{eq:plimit} \& \eqref{eq:elimit} can be parameterised using def. \ref{def:ULflex} and expressed as:
\begin{align}
    \mathcal{F}^{EV}(\underline{p},\overline{p}, \underline{e}, \overline{e}) = \mathcal{F}\left(\vec{u}^{EV}, \vec{l}^{EV}\right) \label{eq:ev_equivalance}
\end{align}
\begin{align}
    where&\nonumber\\
    &u_k^{EV}=\min\left[k\overline{p}\Delta\text{t},\overline{e}-(T-k)\underline{p}\Delta\text{t}\right], &&\: k\in \mathcal{T} \label{ev_u_k}\\
    &l_k^{EV}=\max\left[k\underline{p}\Delta\text{t},\underline{e}-(T-k)\overline{p}\Delta\text{t}\right], &&\: k\in \mathcal{T} \label{l_ev_k}
\end{align}
\end{theorem}
\begin{proof}
We note that the EV charging constraints \eqref{eq:plimit}-\eqref{eq:elimit} directly map onto $u_1=\overline{p}$, $l_1=\underline{p}$, $u_T=\overline{e}$ and $l_T=\underline{e}$. The other components of $\vec{u}$ and $\vec{l}$ are defined as necessary consequences of the properties in Definition~\ref{def:ULflex} and do not further constrain the feasible space.

Concavity of ${}^0\vec{u}$ with $u_1=\overline{p}$ implies $\vec{u}_k\le k \overline{p} \Delta\text{t}$, and convexity of ${}^0\vec{l}$ with $l_1=\underline{p}$ implies $\vec{l}_k\ge k \underline{p} \Delta\text{t}$. Property (iii) results in additional coupling constraints, yielding
    \begin{align}
        u_k &= \min\left[u_T-l_{T-k}, k\overline{p}\Delta\text{t}\right],\quad &&k\in\mathcal{T}\label{eq:u_n1}\\
        l_k &= \max\left[l_T-u_{T-k}, k\underline{p}\Delta\text{t}\right],\quad &&k\in\mathcal{T}\label{eq:l_n1}\
    \end{align}
Using \eqref{eq:l_n1} to replace $l_{T-k}$ in \eqref{eq:u_n1} and $u_T = \overline{e}$, $l_T = \underline{e}$:
\begin{align}
    u_k &= \min\left[k \overline{p}\Delta\text{t}, \overline{e}-\max\left[\underline{e}-u_k, (T-k)\underline{p}\Delta\text{t}\right]\right]\nonumber\\
    &=\min\left[k \overline{p}\Delta\text{t},\overline{e}-\underline{e}+u_k, \overline{e}-(T-k)\underline{p}\Delta\text{t}\right]\nonumber\\
    &=\min\left[k \overline{p}\Delta\text{t}, \overline{e}-(T-k)\underline{p}\Delta\text{t}\right],
\end{align}
where the final step follows from the fact that $\overline{e}-\underline{e}+u_k\geq u_k$, proving \eqref{ev_u_k}. The derivation of \eqref{l_ev_k} follows similarly. It is readily verified that these parameter vectors satisfy all three properties in Definition~\ref{def:ULflex}.\par
\end{proof}

After representing EVs using the UL parameters, they can be aggregated using Theorem \ref{theorem:ul-aggregation} to get their aggregate flexibility as follows.
\begin{corollary}[Aggregate UL representation of EVs]\label{lemma: agg_ev}
The feasible set of attainable aggregated flexibility for a fleet of $N$ EVs can be computed by Minkowski summation of individual UL-feasible sets for each EV, which is additive in parameters, i.e.
\begin{align}
    \mathcal{F}(\Vec{u}_1^{EV},\Vec{l}_1^{EV})\oplus\cdots \oplus &\mathcal{F}(\Vec{u}_n^{EV},\Vec{l}_n^{EV}) \nonumber\\
    &= \mathcal{F}\left(\sum_{n\in \mathcal{N}}\Vec{u}_{n}^{EV},\sum_n\Vec{l}_n^{EV}\right) \label{eq:Minkowski-EV} 
\end{align}
\end{corollary}

A single EV's flexibility region represented by  $\mathcal{F}^{EV}(\underline{p},\overline{p}, \underline{e}, \overline{e})$  contains all permissible power signals by which it can charge to its desired level without violating the given parameters as given in Definition \ref{def_1}. Theorem \ref{theorem: EV's UL flexibility} gives the exact equivalence with a UL-flexibility representation for a single EV. Therefore, the exact feasible flexibility of a fleet of EVs is given by the Minkowski summation of their UL representations, which is a trivial operation, as shown in Corollary \ref{lemma: agg_ev}.\par
We note that the number of constraints in the polytope representations can be drastically reduced in special cases. For example, if full charging is required ($\underline{e}=\overline{e})$ or if vehicles have no minimum charge and no minimum charging power level ($\underline{e}=\underline{p}=0$, so that $\vec{l}=0$).

\begin{figure*}
    \centering
    \includegraphics[width=0.8\linewidth]{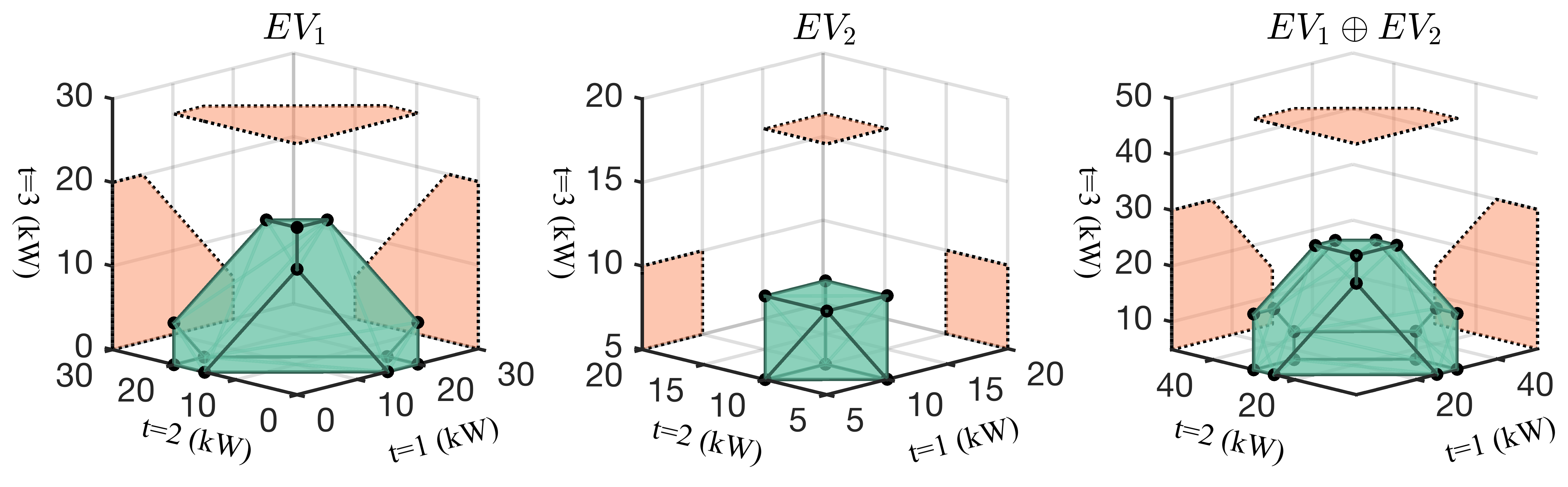}
    \caption{Graphical representation of 3-d polytope representing the feasible flexibility of EV 1, EV 2 and EV 1 $\oplus$ EV 2 respectively as described in Section \ref{section: motivating example}.}
    \label{fig:msumviz}
\end{figure*}
\section{Numerical examples}
We consider the same set of EVs as detailed in Section \ref{section: motivating example}. The UL parameters for both EVs are 
\begin{align}
    &\Vec{u}_1 = [ 20, 25, 25] \: \:  &&\Vec{l}_1 = [ 0, 5, 15], \label{eq:ul_example1} \\
    & \Vec{u}_2 = [ 10, 20, 30] \: \: &&\Vec{l}_2 = [ 5, 10, 20].  \label{eq:ul_example2}
\end{align}
Using Lemma \ref{lemma: agg_ev}, the aggregated UL parameters for the two EVs are
\begin{align}
    \Vec{u}^{tot} &= \sum_{n={1,2}}\Vec{u}_n = [30,45,55],\label{eq:ul_example3}\\
    \Vec{l}^{tot} &= \sum_{n={1,2}}\Vec{l}_n = [5,15,35].\label{eq:ul_example4}
\end{align}
The H-space representation of the polytope representing the aggregate flexibility of the two EVs is given by 
\begin{equation} \label{eq:ul_example5}
\mathcal{P}(\vec{u}^{tot},\vec{l}^{tot}) = \left\{ \vec{p} \in \mathbb{R}^3 \Bigg| \begin{bmatrix}
    \:\:\: A\\
    -A
\end{bmatrix}\vec{p} \preceq 
\begin{bmatrix}
    \:\:\:B\ \vec{u}^{tot}\\
    -B\ \vec{l}^{tot}
\end{bmatrix},
 \right\}
 \end{equation}
where
\begin{equation}
 A = \left[\begin{array}{*{3}c}
1 & 0 & 0 \\
0 & 1 & 0 \\
0 & 0 & 1 \\
1 & 1 & 0 \\
1 & 0 & 1 \\
0 & 1 & 1 \\
1 & 1 & 1 \\
\end{array}\right], \quad 
B = \left[\begin{array}{*{3}c}
1 & 0 & 0 \\
1 & 0 & 0 \\
1 & 0 & 0 \\
0 & 1 & 0 \\
0 & 1 & 0 \\
0 & 1 & 0 \\
0 & 0 & 1 \\
\end{array}\right].
 \nonumber
\end{equation}
The polytopes representing the feasible flexibility region of $EV_1$, $EV_2$ and $EV_1 \oplus EV_2$ are plotted in \figref{fig:msumviz} for three-time steps.  

\section{Computational efficiency}
To assess the effect of scaling on computational times and peak memory usage for the proposed UL-flexibility representation, a simple linear problem is considered. The objective of the LP is to find the maximum aggregate capacity that an EV fleet could deliver for the considered time window. This is achieved by  solving
\begin{align}
    \max {\tau} \label{case_study1}
\end{align}
subject to, for the direct aggregation case:
\begin{subequations}
\begin{align}
    &\underline{p}_n \leq p_{n,t} \leq \bar{p}_n &&\forall\: n \in \mathcal{N};t\in\mathcal{T}\label{case_study2}\\
    &\underline{e}\leq \sum_{k=1}^t p_{n,k}\Delta t \leq \bar{e}_n &&\forall\: n \in \mathcal{N}; t \in \mathcal{T}\label{case_study3}\\
    &\sum_{n\in \mathcal{N}}p_{n,t} \geq \tau && \forall \: t \in \mathcal{T}\label{case_study5}
\end{align}
\end{subequations}
or for the UL-flexibility case:
\begin{subequations}
\begin{align}
    \vec{p}^{agg} &\succeq \tau \\
       \begin{bmatrix}
    \:\:\: A\\
    -A
        \end{bmatrix}\vec{p}^{agg} &\preceq 
        \begin{bmatrix}
    \:\:\:B\vec{u}^{tot}\\
    -B\vec{l}^{tot}
        \end{bmatrix}\\
        \vec{u}^{tot}=\sum_n \vec{u}_n &\text{ \& } \vec{l}^{tot}=\sum_n \vec{l}_n \quad \forall \: n \in \mathcal{N}
\end{align}
\end{subequations}
where $A$ \& $B$ are the matrices defined by \eqref{perm_a_b}. $\vec{u}_n$ and $\vec{l}_n$ are generated for each individual EVs using \eqref{ev_u_k} and \eqref{l_ev_k}.

The above problem was solved for a range of vehicle counts and time steps. Real-world transactions from charging stations in the Netherlands were used and applied to maximise the charging demand between 6 pm and 7 pm. Individual charging transactions were preprocessed to consider computing the minimum and maximum charged energy during the one-hour window, given the arrival and departure times, maximum charging power and the need to complete the charging transaction by the departure time. Transactions that did not (or not fully) cover the 6-7 pm window were discarded. The scaling with time steps was investigated by dividing the considered 1-hour window into $T\in\{1,2,4,16\}$  intervals, resulting in timesteps of sizes 1, 0.5, 0.25 and 0.0625 hours. The simulation was run for different fleet sizes, consisting of  $N\in\{121, 5700, 36108, 90084, 166286,245706\}$ EVs.  The computational complexity of the solved problem is indicative of general optimization problems using UL-flexibility for EV aggregation.\par

The optimization problem was solved using a machine configuration featuring the Apple M2 MAX chip with a 12-core CPU, macOS Ventura Version 13.5.1, 32GB RAM, in conjunction with Python 3.10.11 and the Gurobi 10.0.2 optimization solver. 
The code used in this work is available at \cite{nanda_kishor_panda_2024_10817943} 
\par

\begin{figure*}
    \centering
    \includegraphics[width=1\linewidth]{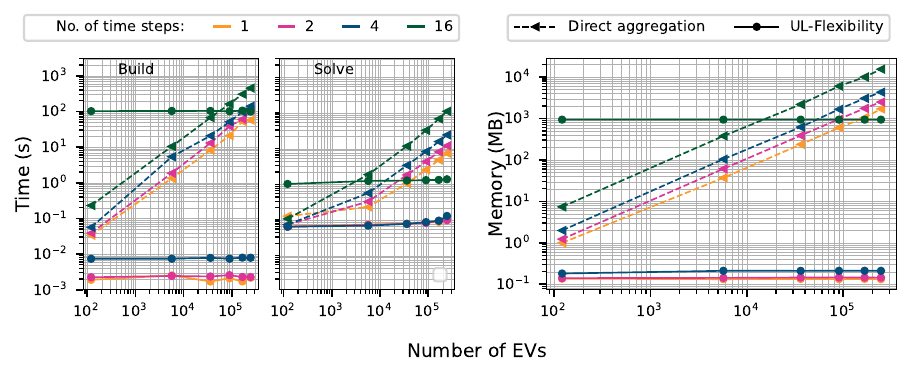}
    \caption{Comparison of computational resources used by the UL-flexibility approach with direct aggregation for different time steps and quantities of EVs. The results of the base case are shown using dashed lines.}
    \label{fig:time_memory}
\end{figure*}

Solve and build time, along with the peak memory use, was recorded and plotted in \figref{fig:time_memory}. As expected from the formulation and supported by the results, the direct aggregation approach's computational time and memory requirements increase as a function of the number of time intervals $T$ and vehicles $N$. In contrast, the UL-flexibility approach has requirements independent of $N$. Both time and memory requirements are minimal for a low number of time intervals $T$ (up to 4, in this case), but they increase sharply for larger numbers of time intervals. Nevertheless, for sufficiently large $N$, the UL-flexibility approach is always more efficient. 

\section{Conclusion}
As shown in earlier sections, the proposed UL-flexibility can be easily used to aggregate feasible flexibility exactly using set-based aggregation, replacing an NP-hard operation. Unlike direct aggregation, the complexity of the proposed approach is independent of the number of assets to aggregate. Moreover, UL-flexibility improves on approximate flexibility aggregation algorithms by its exact aggregation properties, which is beneficial for aggregators in planning and operational decision-making. Hence, the proposed methodology is an efficient and scalable approach for aggregating EVs, especially when optimizing their charging behaviour across short time windows. In future work, populations with varying and potentially stochastic arrival and departure times will be considered.

\balance
\bibliographystyle{IEEEtran}
\bibliography{reference}

\begin{thebibliography}{10}
\providecommand{\url}[1]{#1}
\csname url@samestyle\endcsname
\providecommand{\newblock}{\relax}
\providecommand{\bibinfo}[2]{#2}
\providecommand{\BIBentrySTDinterwordspacing}{\spaceskip=0pt\relax}
\providecommand{\BIBentryALTinterwordstretchfactor}{4}
\providecommand{\BIBentryALTinterwordspacing}{\spaceskip=\fontdimen2\font plus
\BIBentryALTinterwordstretchfactor\fontdimen3\font minus
  \fontdimen4\font\relax}
\providecommand{\BIBforeignlanguage}[2]{{%
\expandafter\ifx\csname l@#1\endcsname\relax
\typeout{** WARNING: IEEEtran.bst: No hyphenation pattern has been}%
\typeout{** loaded for the language `#1'. Using the pattern for}%
\typeout{** the default language instead.}%
\else
\language=\csname l@#1\endcsname
\fi
#2}}
\providecommand{\BIBdecl}{\relax}
\BIBdecl

\bibitem{Muratori2018}
M.~Muratori, ``Impact of uncoordinated plug-in electric vehicle charging on
  residential power demand,'' \emph{Nature Energy}, vol.~3, no.~3, pp.
  193--201, Jan. 2018.

\bibitem{Smart2015}
J.~G. Smart and S.~D. Salisbury, ``Plugged in: How americans charge their
  electric vehicles,'' Tech. Rep., Jul. 2015.

\bibitem{taha2023efficient}
F.~A. Taha, T.~Vincent, and E.~Bitar, ``An efficient method for quantifying the
  aggregate flexibility of plug-in electric vehicle populations,'' 2023.

\bibitem{20.500.11850/202028}
F.~L. Müller, G.~V. Szabó, O.~Sundström, and J.~Lygeros,
  ``\BIBforeignlanguage{en}{Aggregation and disaggregation of energetic
  flexibility from distributed energy resources},''
  \emph{\BIBforeignlanguage{en}{IEEE Transactions on Smart Grid}}, vol.~10,
  no.~2, pp. 1205 -- 1214, 2019-03.

\bibitem{XU2014582}
Z.~Xu, Z.~Hu, Y.~Song, W.~Zhao, and Y.~Zhang, ``Coordination of pevs charging
  across multiple aggregators,'' \emph{Applied Energy}, vol. 136, pp. 582--589,
  2014.

\bibitem{8783654}
Q.~Huang, X.~Wang, J.~Fan, S.~Qi, W.~Zhang, and C.~Zhu, ``V2g optimal
  scheduling of multiple ev aggregator based on tou electricity price,'' in
  \emph{2019 IEEE International Conference on Environment and Electrical
  Engineering and 2019 IEEE Industrial and Commercial Power Systems Europe
  (EEEIC / I\&CPS Europe)}, June 2019, pp. 1--6.

\bibitem{verbist2023impact}
F.~Verbist, N.~K. Panda, P.~P. Vergara, and P.~Palensky, ``Impact of dynamic
  tariffs for smart ev charging on lv distribution network operation,'' 2023.

\bibitem{Tiwary_2008}
H.~R. Tiwary, ``On the hardness of computing intersection, union and~minkowski
  sum of polytopes,'' \emph{Discrete {\&} Computational Geometry}, vol.~40,
  no.~3, pp. 469--479, jul 2008.

\bibitem{homothet_zhao}
L.~Zhao, W.~Zhang, H.~Hao, and K.~Kalsi, ``A geometric approach to aggregate
  flexibility modeling of thermostatically controlled loads,'' \emph{IEEE
  Transactions on Power Systems}, vol.~32, no.~6, pp. 4721--4731, 2017.

\bibitem{7403253}
F.~L. Müller, O.~Sundström, J.~Szabó, and J.~Lygeros, ``Aggregation of
  energetic flexibility using zonotopes,'' in \emph{2015 54th IEEE Conference
  on Decision and Control (CDC)}, 2015, pp. 6564--6569.

\bibitem{Barot2017}
S.~Barot and J.~A. Taylor, ``A concise, approximate representation of a
  collection of loads described by polytopes,'' \emph{International Journal of
  Electrical Power {\&} Energy Systems}, vol.~84, pp. 55--63, Jan. 2017.

\bibitem{Wu2020}
D.~Wu, P.~Wang, X.~Ma, and K.~Kalsi, ``Scheduling and control of flexible
  building loads for grid services based on a virtual battery model,''
  \emph{{IFAC}-{PapersOnLine}}, vol.~53, no.~2, pp. 13\,333--13\,338, 2020.

\bibitem{ulbig2015analyzing}
A.~Ulbig and G.~Andersson, ``Analyzing operational flexibility of electric
  power systems,'' \emph{International Journal of Electrical Power \& Energy
  Systems}, vol.~72, pp. 155--164, 2015.

\bibitem{Brinkel2022}
N.~Brinkel, J.~Hu, L.~Visser, W.~van Sark, and T.~AlSkaif, ``Scheduling
  electric vehicle fleets as a virtual battery under uncertainty using quantile
  forecasts,'' in \emph{2022 {IEEE} International Conference on Communications,
  Control, and Computing Technologies for Smart Grids ({SmartGridComm})}.\hskip
  1em plus 0.5em minus 0.4em\relax {IEEE}, Oct. 2022.

\bibitem{9069450WANG}
B.~Wang, D.~Zhao, P.~Dehghanian, Y.~Tian, and T.~Hong, ``Aggregated electric
  vehicle load modeling in large-scale electric power systems,'' \emph{IEEE
  Transactions on Industry Applications}, vol.~56, no.~5, pp. 5796--5810, 2020.

\bibitem{9772074_ep}
M.~P. Evans, S.~H. Tindemans, and D.~Angeli, ``Flexibility framework with
  recovery guarantees for aggregated energy storage devices,'' \emph{IEEE
  Transactions on Smart Grid}, vol.~13, no.~5, pp. 3519--3531, 2022.

\bibitem{cruise2019optimal}
J.~Cruise and S.~Zachary, ``Optimal scheduling of energy storage resources,''
  2019.

\bibitem{Zachary2021}
\BIBentryALTinterwordspacing
S.~Zachary, S.~H. Tindemans, M.~P. Evans, J.~R. Cruise, and D.~Angeli,
  ``Scheduling of energy storage,'' \emph{Philosophical Transactions of the
  Royal Society A: Mathematical, Physical and Engineering Sciences}, vol. 379,
  no. 2202, p. 20190435, Jun. 2021. [Online]. Available:
  \url{https://doi.org/10.1098/rsta.2019.0435}
\BIBentrySTDinterwordspacing

\bibitem{mukhiexact}
K.~Mukhi and A.~Abate, ``An exact characterisation of flexibility in
  populations of electric vehicles,'' in \emph{2023 62nd IEEE Conference on
  Decision and Control (CDC)}.\hskip 1em plus 0.5em minus 0.4em\relax IEEE,
  2023, pp. 6582--6587.

\bibitem{sigl2023mrepresentation}
S.~Sigl and M.~Althoff, ``M-representation of polytopes,'' 2023.

\bibitem{Barot_2017}
S.~Barot, ``Aggregate load modeling for demand response via the minkowski
  sum,'' Ph.D. dissertation, University of Toronto, 2017.

\bibitem{Grünbaum_Klee_Ziegler_2003}
B.~Grünbaum, V.~Klee, and G.~M. Ziegler, \emph{Polytopes, Definition and
  FundamentalProperties of Polytopes}, second edition~ed., ser. Graduate Texts
  in Mathematics.\hskip 1em plus 0.5em minus 0.4em\relax Springer
  Science+Business Media, LLC, 2003, p. 31–52.

\bibitem{nanda_kishor_panda_2024_10817943}
\BIBentryALTinterwordspacing
N.~K. Panda, ``nkpanda97/ul-flexibility: Version 2023.01,'' Mar. 2024.
  [Online]. Available: \url{https://zenodo.org/doi/10.5281/zenodo.10811009}
\BIBentrySTDinterwordspacing

\end{thebibliography}
\end{document}